\documentclass[USenglish,twocolumn]{article}
\usepackage{amsmath,amssymb,amsfonts,bbm,graphicx,color}
\usepackage{physics}

\usepackage[utf8]{inputenc}%(only for the pdftex engine)
\usepackage[big]{dgruyter}

\newcommand{\Hq}{\mathbf{H}_q}
\newcommand{\nmdiv}{\mathcal{N}_\text{\tiny{M}}}
\newcommand{\nmblp}{\mathcal{N}_\text{\tiny{BLP}}}

\begin{document}
  \articletype{Research Article{\hfill}Open Access}

\author*[1]{Matteo A. C. Rossi}
\author[2]{Marco Cattaneo} 
\author[2]{Matteo G. A. Paris} 
\author[3]{Sabrina Maniscalco}
\affil[1]{QTF Centre of Excellence, Turku Centre for Quantum Physics, Department
of Physics and Astronomy, University of Turku, FI-20014 Turun Yliopisto,
Finland, E-mail: matteo.rossi@utu.fi} \affil[2]{Quantum Technology Lab,
Dipartimento di Fisica ``Aldo Pontremoli'', Universit\`a degli Studi di Milano,
I-20133 Milano, Italy} \affil[3]{QTF Centre of Excellence, Turku Centre for
Quantum Physics, Department of Physics and Astronomy, University of Turku,
FI-20014 Turun Yliopisto, Finland\\
QTF Centre of Excellence, Department of Applied Physics, School of Science,
Aalto University, FI-00076 Aalto, Finland}

\title{\LARGE Non-Markovianity is not a resource for quantum spatial search on a
star graph subject to generalized percolation}
\runningauthor{M. A. C. Rossi et al.}
\runningtitle{Non-Markovianity is not a resource for quantum spatial search...}

  %\subtitle{...}
\begin{abstract}
{Continuous-time quantum walks may be exploited to enhance spatial search, i.e.,
for finding a marked element in a database structured as a complex network.
However, in practical implementations, the environmental noise has detrimental
effects, and a question arises on whether noise engineering may be helpful in
mitigating those effects on the performance of the quantum algorithm. Here we
study whether time-correlated noise inducing non-Markovianity may represent a
resource for quantum search. In particular, we consider quantum search on a star
graph, which has been proven to be optimal in the noiseless case, and analyze
the effects of independent random telegraph noise (RTN) disturbing each link of
the graph. Upon exploiting an exact code for the noisy dynamics, we evaluate
the quantum non-Markovianity of the evolution, and show that it cannot be
considered as a resource for this algorithm, since its presence is correlated
with lower probabilities of success of the search.}
\end{abstract}
  %\keywords{keywords, keywords}
  \journalname{QMTR}

  \DOI{DOI}
  \startpage{1}
  \received{..}
  \revised{..}
  \accepted{..}

  \journalyear{????}
  \journalvolume{?}
  \journalissue{??}

\maketitle
\section{Introduction}
There is a close connection between quantum metrological precision bounds and
quantum computation speed-up limits, e.g. the search time in a database
\cite{DemkowiczDobrzaski2015}. In turn, the interest in quantum computation
relies on its ability to outperform standard classical computation in solving
some peculiar tasks. Among these is the problem of finding a certain element
with a  given property in a disordered database of $N$ items. Grover's quantum
algorithm \cite{Grover1996} retrieves the specified target at time of order
$T={O}(\sqrt{N})$ instead of the classical $T={O}(N)$. Moreover, this quantum
speed-up has been proven to be optimal \cite{Zalka1999}. Quantum spatial search
\cite{Aaronson2003} is the generalization of this problem to a database
characterized by a complex structure, i.e., a database whose elements are
distributed in space and connected by links according to a certain topology.
Such a database can be described by a graph.
\par  
Different methods of solving the problem of quantum spatial search have been
proposed
\cite{Aaronson2003,Childs2004,ambainis2005coins,Tulsi2008,Novo2018}.
Here we focus on the algorithm based on continuous-time quantum walks (CTQWs)
\cite{Farhi1998trees}, introduced by Childs and Goldstone \cite{Childs2004},
that can achieve the optimal speed-up $T={O}(\sqrt{N})$ on certain topologies,
such as the complete graph or the hypercube. Many other graphs are suitable for
quantum spatial search using this algorithm: For instance, the star graph was
recently proven to be optimal \cite{Cattaneo2018}. However, despite a great
theoretical effort in characterizing the networks that are suitable for the
search and to find more efficient versions of the algorithm, only few studies
address the effects of noise on quantum spatial search via CTQWs.
\par
The presence of broken links in complex networks has been investigated in
\cite{Novo2015}, and it has been shown that the coupling of the system to a
thermal bath may improve the performance of the algorithm affected by static
disorder \cite{Novo2017}. Changing the complex structure of the graph after a
time interval $\tau$, i.e., creating random temporal networks, can lead to a
dynamical topology suitable for search as well \cite{Chakraborty2017}, while the
first study of a fully-dynamical description of the noise has been recently
presented \cite{Cattaneo2018}, in particular introducing classical random
telegraph noise (RTN) affecting the hopping rate of the links of the network.
The effect of RTN on the dynamics of quantum systems has been widely studied in
the literature, being a typical model for noise affecting solid state devices
\cite{Galperin2006,Abel2008,Cheng2008} and used as a building block of $1/f$
noise \cite{Paladino2002,Paladino2014}. Many works have focused on one- or
two-qubit systems \cite{Galperin2006,Benedetti2013a,Rossi2016,Cialdi2017}, with
studies of its effect on CTQWs appearing in recent literature
\cite{Benedetti2016,Siloi2017a,Rossi2017a,Benedetti2019}.
\par
A key concept in the field of open quantum systems is non-Markovianity.
Depending on different points of view, it expresses the divisibility of the
quantum map describing the evolution of the system \cite{Rivas2014} or the
backflow of quantum information going from the environment to the system
\cite{Breuer2009}. A crucial point in the study of non-Markovianity relies on
understanding when its presence is a resource, i.e., when non-Markovianity
enhances the results of the particular task we want to achieve using the quantum
system. Quantum non-Markovianity was proven to be a resource in different
scenarios for quantum information processing \cite{Huelga2012,Bylicka2014},
teleportation \cite{Laine2014}, computation \cite{dong2018}, metrology
\cite{Chin2012}. Systems affected by RTN can exhibit either Markovian and
non-Markovian quantum dynamics, depending on the parameters and on the type of
interaction with the environment \cite{Benedetti2014c,Rossi2016,Rossi2017a}: the
latter has been shown to allow for recoherence effects in qubit systems
\cite{Benedetti2014c}, while it induces localization in quantum walks on
lattices \cite{Benedetti2016,Rossi2017a}.
\par
In this paper, we address the role of quantum non-Markovianity in the
computational task of quantum spatial search. We answer the questions: is
non-Markovianity a resource for quantum spatial search via CTQW? Does its
presence improve the performance of the algorithm?
\par
As a matter of fact, the dynamics of the CTQW on graphs subject to dynamical
noise is obtained by Montecarlo simulation of the noise \cite{Cattaneo2018}.
However, the study of non-Markovianity requires higher precision and numerical
stability, therefore in this paper we employ a numerically exact technique to
obtain the state of the walker at a generic time $t$. This technique, valid for
any system subject to classical dynamical noise, was first proposed in
\cite{Joynt2011} and specifically used to study the dynamics of small quantum
systems, such as one or two qubits perturbed by random telegraph noise
\cite{Cheng2008,Rossi2016}. Here, we develop a fast code that allows us to scale
up the technique to larger quantum systems. We discuss the general technique in
Sec.~\ref{sec:method}, while the code we used to implement it is available on
GitHub \cite{QuasiHamiltonianRTN}.
\par
The paper is structured as follows: in Sec.~\ref{sec:algorithm} we review the
quantum spatial search algorithm based on CTQW and we discuss the noise model.
In Sec.~\ref{sec:nonM} we review the concept of quantum non-Markovianity and
introduce the measures we employ to study the noisy evolution. In
Sec.~\ref{sec:method} we present the analytical method we have used to calculate
the evolution of the quantum walk subject to dynamical noise. In
Sec.~\ref{sec:results} we discuss the results, while
Sec.~\ref{sec:concludingRemarks} closes the paper with some concluding remarks.
%%%
\section{Noisy quantum spatial search}
\label{sec:algorithm}
We model our structured database as a given graph $G$ composed of $N$ nodes, and
we want to find the marked element $w$, called target node. Any graph is
characterized by an adjacency matrix $A$, whose elements are defined as
\begin{equation}
A_{ij} = \begin{cases}
  1 & \text{if nodes $i,j$ connected} \\
  0 & \text{otherwise}.
\end{cases}
\end{equation}

We want to run a CTQW on this graph in order to find $w$. The Hilbert space of
the walker is $\mathcal{H}=\text{span}\{\ket{j}\}$ with $j=1,\ldots,N$, where
$\ket{j}$ is the single-particle localized state associated to the node $j$.
According to the original definition \cite{Farhi1998trees}, the Hamiltonian of
the walk is proportional to the Laplacian matrix of the graph $L$, defined as $L
= D - A$, where $D$ is the degree matrix, a diagonal matrix such that $D_{jj}$
is the number of links connected to node $j$. To perform the spatial search, we
add to the original Hamiltonian a projector onto the target node, in order to
localize the walker there. Therefore, the Hamiltonian of the algorithm reads
\begin{equation}
\label{eqn:Hamiltonian}
H=\gamma L + H_w = \gamma L-\dyad{w},
\end{equation}
where $H_w=-\dyad{w}$ is called \textit{oracle Hamiltonian}, $\gamma$ is a
suitable coupling constant and $L$ is the Laplacian matrix associated to $G$.

The initial state of the quantum walk is the fully delocalized state $\ket{s}$:
\begin{equation}
\label{eqn:initialState}
\ket{s}=\frac{1}{\sqrt{N}}\sum_{j=1}^N \ket{j},
\end{equation}
and the state at time $t$ reads
\begin{equation}
\ket{\psi(t)}=e^{-iHt}\ket{s}.
\end{equation}
If, at time $t$, we measure the walker in the node basis, the probability of
obtaining the target node is given by $  p(t) = \abs{\braket{w}{\psi(t)}}^2$. We
assume that we can choose to measure at the time $T$ for which the above
probability is maximal, and we define the success probability of the algorithm
as
\begin{equation}
     p_\text{\tiny{succ}} = \abs{\braket{w}{\psi(T)}}^2
\end{equation}
We want to maximize $p_\text{\tiny{succ}}$ keeping $T$ as short as possible. The
algorithm is optimal on the given graph $G$ if there exists a time
$T=O(\sqrt{N})$ and a suitable constant $\gamma$ for which the probability of
success is close to $1$.

\begin{figure}
  \centering
  \includegraphics[width=.7\columnwidth]{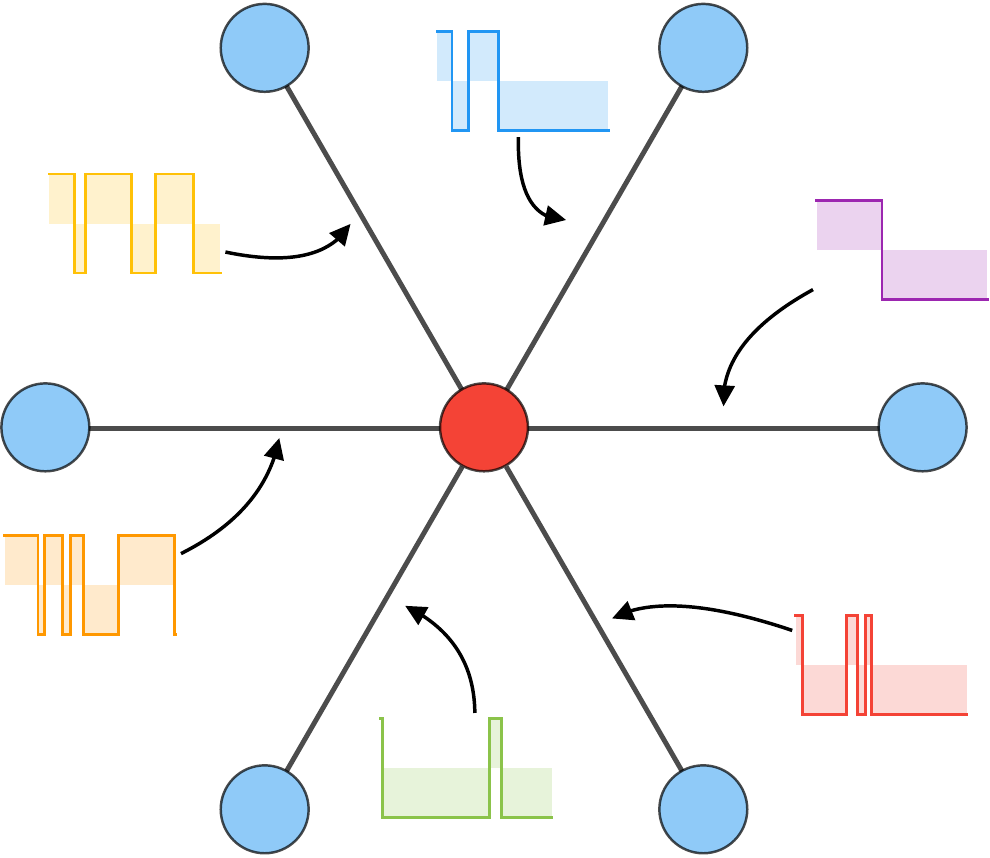}
  \caption{Pictorial representation of the model described in Section 
  \ref{sec:algorithm}: the links between the nodes of a graph (in this work, 
  we focus on the star graph) are affected by independent sources of RTN, all 
  characterized by the same switching rate $\mu$ and noise strength $\nu$. The 
  red node corresponds to the marked node in the Hamiltonian, $\ket{w}$.}
  \label{fig:model}
\end{figure}

We now describe how to introduce dynamical noise on the algorithm, following the
approach of \cite{Cattaneo2018}: a pictorial representation of the model is
shown in Fig.~\ref{fig:model}. We insert independent random telegraph noise
(RTN) perturbing the hopping rate of each link of the graph, where the RTN is a
classical dynamical noise that can assume only two values, say $g(t)=\pm 1$, and
the probability of switching value $n$ times in a time $t$ follows the Poisson
distribution
\begin{equation}
p_\mu(n,t)=e^{-\mu t}\frac{(\mu t)^n}{n!},
\end{equation}
where $\mu$ is called switching rate.

Therefore, RTN describes a stationary stochastic process with autocorrelation
function
\begin{equation}
\langle g(\tau)g(0)\rangle=e^{-2\mu\abs{\tau}},
\end{equation}
corresponding to a Lorentzian spectrum.

CTQWs affected by RTN have been studied in the recent past for one-dimensional
lattices \cite{Benedetti2016,Siloi2017a,Rossi2017a}, and for quantum spatial
search on graphs \cite{Cattaneo2018}. Here we consider independent random
telegraph noise perturbing each link of the complex network with the same
switching rate $\mu$, and we accordingly modify the Laplacian matrix in
Eq.~\eqref{eqn:Hamiltonian} as follows.

The noise is described by the $N\times N$ matrix $\mathbf{g}(t)$, where $N$ is
the number of nodes in the graph and $g_{jk}(t)$ is the stochastic process
describing the noise on the link connecting $j$ to $k$. The matrix
$\mathbf{g}(t)$ is thus symmetric, zero-diagonal and has only $l$ independent
entries, where $l$ is the number of links in the graph. Keeping in mind that the
noise realizations on different links are uncorrelated, we have the following
autocorrelation function, for the non-zero elements of $\mathbf{g}(t)$
\begin{equation}
\langle g_{jk}(\tau)g_{j'k'}(0)\rangle=e^{-2\mu\abs{\tau}}
(\delta_{jj'}\delta_{kk'}+\delta_{jk'}\delta_{kj'})\,.
\end{equation}

The noisy Laplacian $L^{(\mathbf{g})}(t)$ thus reads
\begin{equation}
  \label{eqn:noisyHam}
  L^{(\mathbf{g})}_{jk}(t)=\begin{cases}
  -\left[1+\nu g_{jk}(t)\right]&\textnormal{ if }(j,k)\textrm{ connected}\\
  D_{jk}+\nu \sum_{i=1}^Ng_{ik}(t)&\textnormal{ if }j=k \\
  0&\textnormal{ otherwise}
  \end{cases}
  \end{equation}
where $\nu \in [0, 1]$ is the relative noise strength, assumed to be the same
for all the links. The Hamiltonian of the noisy walk, replacing the one in
Eq.~\eqref{eqn:Hamiltonian}, is now a function of the stochastic process
$\mathbf{g}(t)$ and reads
\begin{equation}
  H^{(\mathbf{g})}(t) = \gamma L^{(\mathbf{g})}(t) - \ket{w}\!\!\bra{w}\,.
\end{equation}

Using the language of open quantum systems, we describe the state of the system
at time $t$ as a density matrix $\rho(t)$. Starting from the initial state
$\rho_0=\ket{s}\!\!\bra{s}$,
\begin{equation}
\label{eqn:evolution}
\rho(t)=\langle U[\mathbf{g}(t)]\rho_0 U^\dag[\mathbf{g}(t)]\rangle_{\{\mathbf{g}(t)\}},
\end{equation}
where $\langle \ldots\rangle_{\{\mathbf{g}(t)\}}$ denotes the average over all
possible realizations of the stochastic process $\mathbf{g}(t)$, while
$U[\mathbf{g}(t)]$ is the unitary evolution operator that drives the evolution
associated to a particular realization of the noise, given by
\begin{equation}
\label{eqn:unOp}
U[\mathbf{g}(t)] =\mathcal{T}\exp{-i\int_0^t ds\,H^{(\mathbf{g})}(s)},
\end{equation}
where $\mathcal{T}$ is the time-ordering operator.

Equation \eqref{eqn:evolution} describes a quantum map sending a density matrix
into a density matrix. Considering the initial time $t_0=0$, for each time $t$
we denote such a map as $\mathcal{E}(t,0)$, defined as
\begin{equation}
\label{eqn:quantumMap}
\mathcal{E}(t,0)\rho_0=\rho(t).
\end{equation}

%%%%
\section{Measures of non-Markovianity}
\label{sec:nonM}
Extending the concept of non-Markovianity for stochastic processes to the
quantum world is not trivial, since the classical definition of Markovianity is
based on probability distributions evaluated at different times, while in
quantum mechanics measuring the state of the system affects it, thus it is not
meaningful anymore to define a general quantity like quantum non-Markovianity
using classical objects such as probability distributions
\cite{Rivas2014,Breuer2016}.

A very well-known class of \textit{master equations}, i.e., evolution equations
for quantum states, is the one written in the
\textit{Gorini-Kossakowski-Sudarshan-Lindblad} (GKSL) form
\cite{Gorini1976,Lindblad1976}. The quantum maps described by the GKSL master
equation (or, more generally, its time-local generalization) are said to be
Markovian because of their \textit{divisibility} property \cite{Rivas2014}:
if, considering also time-inhomogeneous processes,
$\mathcal{E}(t_2,t_1)$ is the quantum map generating the evolution of a quantum
state from $t_1$ to $t_2$, and if this evolution follows the GKSL master
equation, then the quantum map is \textit{divisible} in the sense that
\begin{equation}
\label{eqn:divisibility}
\mathcal{E}(t_3,t_1)=\mathcal{E}(t_3,t_2)\mathcal{E}(t_2,t_1)\quad\forall\,t_1<t_2<t_3.
\end{equation}

This property can be seen as a sort of quantum analogue of the
Chapman-Kolmogorov equation characterizing a Markovian stochastic process.
Furthermore, the GKSL master equation is obtained by imposing some
approximations upon the coupling between system and environment
\cite{Breuer2002}. In particular, weak coupling, Born approximation, and fast
decay of the environment's correlation functions (compared to the typical
time-scale of the evolution of the quantum state) are required. These conditions
can be seen as reflecting a memoryless evolution of the system, thus
strengthening the idea of ``quantum Markovianity'' of the quantum map. We refer
the reader to a standard textbook for a more rigorous explanation of the GKSL
master equation \cite{Breuer2002}.

Further definitions of quantum non-Markovianity have been proposed and used. In
particular, a really common definition is the one based on the \textit{backflow
of quantum information} between system and environment \cite{Breuer2009}. The
choice of a definition rather than another one depends on the specific purposes
for which we want to evaluate quantum non-Markovianity. The main results and
proposals on the topic are reviewed in \cite{Rivas2014,Breuer2016,DeVega2017}.
Moreover, it is known that different definitions follow a hierarchy, i.e., some
classes of definitions are contained in other ones; this aspect, first
discovered in \cite{Chruscinski2014}, has been deeply investigated in a very
recent paper \cite{Li2018}. In addition to the detection of quantum
non-Markovianity of a quantum process, we would like to quantify the amount of
non-Markovianity of a quantum map. Various measures of non-Markovianity have
been introduced in the literature to achieve this goal. In what follows we
explore two measures of quantum non-Markovianity that we will use in our work,
chosen for their significance in the literature and the possibility to compute
them with the problem at hand.

\subsection{Divisibility measure}
The first measure we analyze is strictly related to the definition of
non-Markovianity based on the divisibility of the quantum map. We will employ a
variation of the one proposed in \cite{Hou2015}, which has already been used in
\cite{Benedetti2016} for quantum walks on lattices. 

Suppose that $\mathcal{E}$ is the quantum map describing the evolution of a
quantum state starting at $t=0$, and suppose to take $\rho_0$ as the initial
state. We evaluate the quantity
\begin{equation}
\label{eqn:nonMgammaTauTau1}
\Gamma(\tau,\tau_1)=D(\mathcal{E}(\tau,0)\rho_0,\; \mathcal{E}(\tau,\tau_1)\mathcal{E}(\tau_1,0)\rho_0),
\end{equation}
where $0 \leq \tau_1 \leq \tau$, and $D$ is the trace distance between two
states, defined as:
\begin{equation}
\label{eqn:traceDist}
D(\rho_1,\rho_2)=\frac{1}{2}\abs{\rho_1-\rho_2},
\end{equation}
with $\abs{A}=\Tr\sqrt{A^\dagger A}$ for a square matrix $A$.

Obviously, in the case of time-homogeneous processes,
$\mathcal{E}(\tau,\tau_1)=\mathcal{E}(\tau-\tau_1)$.
Eq.~\eqref{eqn:nonMgammaTauTau1} is basically evaluating how distant the final
state obtained through the complete evolution is, compared to the one for which
the evolution has been stopped and restarted at a certain time $t_1$; it is thus
detecting how $\mathcal{E}$ deviates from divisibility. $\Gamma(\tau,\tau_1)$ is
clearly zero for any $\tau$ and $\tau_1$ if $\mathcal{E}$ is described with a
master equation in the GKSL form.

In order to get a number quantifying the deviation from divisibility, one takes
the maximal deviation from the property of divisible quantum map, i.e., the
maximum over all $\tau$ and $\tau_1$ up to infinity. Therefore, the measure of
non-Markovianity that we employ is
\begin{equation}
\label{eqn:nonMgammaMax}
\nmdiv=\max\limits_{\tau,\tau_1}\Gamma(\tau,\tau_1).
\end{equation}

It is evident that Eq.~\eqref{eqn:nonMgammaMax} does not define a measure of the
non-Markovianity of the quantum map, but only of the evolution of a particular
initial state. Indeed, in \cite{Hou2015} the trace distance in
Eq.~\eqref{eqn:nonMgammaTauTau1} is replaced with a distance in the quantum maps
space. However, in the case at hand the initial state of the system is fixed by
the prescription of the spatial search algorithm, thus
Eq.~\eqref{eqn:nonMgammaMax} is both easier to calculate and appropriate for our
purposes.

\subsection{BLP measure}
Probably the most famous measure of non-Markovianity is the \textit{BLP measure}
\cite{Breuer2009}, based on the backflow of quantum information between system
and environment.

The trace distance between two states is contractive under the action of a
quantum channel, i.e. a completely positive and trace preserving map
\cite{NielsenChuang}. It is straightforward to prove \cite{Breuer2009} that, if
$\mathcal{E}$ is a divisible quantum map Eq.~\eqref{eqn:divisibility}, then the
trace distance of two evolved states (with initial state $\rho_1(0)$ and
$\rho_2(0)$) is monotonically decreasing in time: namely, if
$\rho_1(t)=\mathcal{E}(t,0)\rho_1(0)$ and $\rho_2(t)=\mathcal{E}(t,0)\rho_2(0)$,
\begin{equation}
\label{eqn:monDecTraceDist}
D(\rho_1(t+\tau),\rho_2(t+\tau))\leq D(\rho_1(t),\rho_2(t))\quad\forall \,t,\tau >0.
\end{equation}
This may not be true anymore if the dynamics is non-Markovian. Therefore, let us
define the quantity
\begin{equation}
\label{eqn:derTraceDist}
\sigma(t,\rho_{1,2}(0))=\frac{d}{dt}D(\rho_1(t),\rho_2(t)).
\end{equation}
If $\sigma(t,\rho_{1,2}(0))$ is positive for certain time intervals, then the
quantum map is non-Markovian and, in particular, during those time intervals we
are observing a backflow of quantum information. Indeed, the trace distance
expresses our ability to distinguish the states $\rho_1$ and $\rho_2$
\cite{NielsenChuang}, therefore when it increases we are acquiring more
\textit{quantum information} about the two states.

The BLP measure is defined by integrating over all the time intervals in which
we are gaining quantum information, i.e., in which Eq.~\eqref{eqn:derTraceDist}
is positive, and then taking the maximum upon all the possible pairs of initial
states:
\begin{equation}
\label{eqn:BLP}
\nmblp(\mathcal{E})=\max_{(\rho_1(0),\rho_2(0))}\int_{\sigma >0} dt\,\sigma(t,\rho_{1,2}(0)).
\end{equation}

Following the hierarchy of non-Markovianity measures, there are some dynamical
maps for which the BLP measure is zero, despite being non-Markovian with respect
to the divisibility definition \cite{Rivas2014}. Nonetheless,
Eq.~\eqref{eqn:BLP} is a true measure of non-Markovianity of a quantum map (and
not only of a specific evolution), and it provides the quantifier of a useful
resource  (the backflow of information).

Due to the maximization upon all the possible initial states, the evaluation of
Eq.~\eqref{eqn:BLP} is, in general, a formidable task, only slightly mitigated
by the fact, proven in \cite{Wissmann2012}, that the states of the optimal pair
must lie on the boundary of the space of the density matrices and must be
orthogonal. Given the problem at hand, we choose to fix $\rho_1(0)$ as the
initial state of the spatial search algorithm, and we optimize over the state
$\rho_2(0)$ only.
%%%

\begin{figure*}[t] \includegraphics{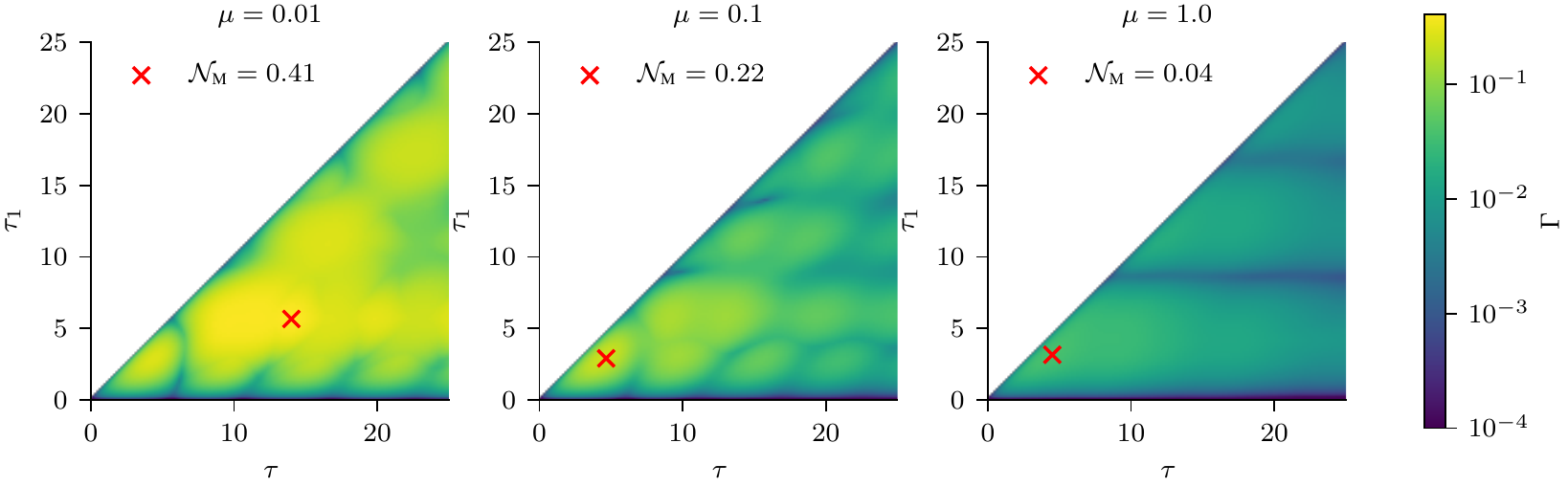}%
\caption{$\Gamma(\tau,\tau_1)$ as a function of $\tau$ and $0 \leq \tau_1 \leq \tau$, for several values of switching rate $\mu$ and noise strength $\nu = 1$, in logarithmic scale. The plots are for the spatial search dynamics on the star graph with $N=7$ and central target node. The red cross marks the maximum value, i.e., the measure $\nmdiv$. We can observe that slow strong noise (deleterious for the algorithm) leads to a higher value of non-Markovianity.}
\label{fig:nonMdivTauTau1}
\end{figure*} 

%%%%
\section{Analytical solution of the noisy dynamics}
\label{sec:method}
The solution of Eq.~\eqref{eqn:evolution} is usually computed numerically,
because of the cumbersome expression that arises in Eq.~\eqref{eqn:unOp} and of
the huge number of possible realizations of the noise. Exact analytical
solutions are possible only in certain cases in which the Hamiltonian commutes
with itself at different times, such as in the case of pure dephasing of qubits
\cite{Benedetti2016}.

Joynt et al. have proposed an exact method of solving the dynamics of a quantum
system coupled to a classical environment modeled as a Markovian stochastic
process, and particularly effective for RTN \cite{Joynt2011,Cheng2008}. The
method allows for analytical results only for a single qubit \cite{Cheng2008},
while it requires numerical matrix diagonalization for higher dimensions,
\cite{Rossi2016}. The strengths of this method are that it gives exact results
up to machine-precision, and it avoids fluctuations typical of Montecarlo
simulations: the drawback, however, is the exponential complexity in terms of
the number of noise fluctuators.

We have implemented the method in Julia \cite{Bezanson2017}, with particular
emphasis on optimization for the problem at hand: the code is available on
GitHub \cite{QuasiHamiltonianRTN}. Based on this code we are able to solve the
dynamics of a CTQW subject to dynamical noise for graphs with up to $N=10$
links. While this number is still quite small, it allows for gaining intuition
on the effects of noise on the spatial search algorithm and the relation to
non-Markovianity.

In this section, we briefly explain how to obtain the exact dynamics of the
system with the method introduced in \cite{Joynt2011}, leaving the full
explanation and proof to the original paper. Suppose to have a $N_q$-dimensional
quantum state, described at time $t$ by the $N_q\times N_q$ density matrix
$\rho(t)$, and a classical system made of $N_c$ states, representing the
possible values of the noise. For example, if the classical noise is a single
fluctuator, $N_c = 2$; if it consists of $N$ independent RTN sources, $N_c =
2^N$.

We start at $t=0$ with $\rho(0)$ and the classical probability distribution
$\mathbf{P}(0)$, describing the initial state of the stochastic process
associated to the classical noise. The Hamiltonian of the quantum system is
$H[g(t)]$, i.e., a function of the stochastic process describing the noise. At
every time instant, to every particular configuration of the noise, which we
label with the index $c \in \{1, \ldots, N_c\}$, corresponds a particular form
of the Hamiltonian $H_c$.

Since we assume a Markovian classical environment, the probability of the
different states is described by the master equation
\begin{equation}
  \frac {d \mathbf{P}(t)} {dt} = \mathbf{V} \mathbf{P}(t),
\end{equation}
where the element $\mathbf{V}_{c,c'}$ of the matrix $\mathbf{V}$ dictates the
transition rates between the states $c$ and $c'$ of the environment. Notice that
$\mathbf{V}$ is time-independent if the stochastic process describing the
environment is homogeneous, as is the case in this work. In the case of a single
RTN with a switching rate $\mu$, we have that $N_c = 2$ and
\begin{equation}
\label{eqn:VrtnMU}
\mathbf{V}_\mu =\begin{pmatrix}
-\mu&\mu\\
\mu&-\mu
\end{pmatrix}.
\end{equation}

For a collection of $N$ independent fluctuators, the matrix $\mathbf{V}$ becomes
\begin{equation}\label{eq:VrtnN}
  \mathbf{V} = \sum_{i=1}^N \mathbf{V}_\mu^{(i)}, \quad \mathbf{V}_\mu^{(i)} = \mathbf{I}_2^{\otimes i-1} \otimes \mathbf{V}_\mu \otimes \mathbf{I}_2^{\otimes N-i}.
\end{equation}

We need to represent the density matrix as a vector, and we do so by employing
the generalized Bloch vector $\mathbf{n}(t)$, a vector of dimension $N_q^2-1$,
with real components
\begin{equation}
  n_i(t) = \frac{\sqrt{N_q}}{2} \Tr \lambda_i \rho(t),
\end{equation}
where $\lambda_j$ are the generators of $SU(N_q)$, and they are $N_q\times N_q$
matrices chosen to satisfy
\begin{equation}
\label{eqn:SUNgenerators}
\Tr \lambda_j =0,\quad \lambda_j^\dagger=\lambda_j,\quad \Tr (\lambda_j\lambda_k) = 2 \delta_{jk}.
\end{equation}
We can go back to the density matrix $\rho(t)$ from the Bloch vector
$\mathbf{n}(t)$ by means of the equation
\begin{equation}
\label{eqn:SUNmixedState}
\rho(t)=\frac{1}{N_q}\left[\mathbf{I}_{N_q}+\sqrt{N_q}\sum_{j=1}^{N_q^2-1} n_j(t)\lambda_j\right],
\end{equation}
where $\mathbf{I}_{N_q}$ denotes the identity in the Hilbert space of the
quantum system.

The action of a unitary operator $U$ onto the density matrix $\rho(t)$ is
translated into the multiplication of the Bloch vector $\mathbf{n}(t)$ by a
transfer matrix $T$ defined as
\begin{equation}
  \mathbf{T}_{ij} = \frac 12 \Tr[\lambda_i U \lambda_j U^\dagger].
\end{equation}

Consider now a short time interval $\Delta t$ in which the environment is in a
fixed state $c$; during $\Delta t$, the unitary evolution is generated by the
Hamiltonian $H_c$: $U_c(\Delta t)=\exp [-iH_c \Delta t]$. The corresponding
transfer matrix $\mathbf{T}_c$ is generated by the matrix
\begin{equation}\label{eq:quasihamiltonian_generators}
  G_c = i \lim_{\Delta t \to 0} \frac{\mathbf{T}_c - \mathbf{I}_{N_q}}{\Delta t} = \frac i 2 \sum_{i,j=1}^{N_q} \Tr \left([\lambda_i, \lambda_j] H_c \right).
\end{equation}

In their paper \cite{Joynt2011}, \citeauthor{Joynt2011} introduced the
\emph{quasi-Hamiltonian} matrix
\begin{equation}
\label{eqn:quasiHamDivided}
\Hq=i\mathbf{V}\otimes\mathbf{I}_{N_q^2-1} + \bigoplus_{i=c}^{N_c} G_c,
\end{equation}
where the second term is a direct sum of all the generators defined in
\eqref{eq:quasihamiltonian_generators}, and showed that the dynamics of the
system, averaged all the possible realizations of the stochastic process
describing the noise (as defined in Eq. \eqref{eqn:unOp}), is given by
\begin{equation}\label{eq:hq_final_equation}
  \mathbf{n}(t) = \mel{\mathbf{1}}{\exp(-i \Hq t)}{p_0}\cdot\mathbf{n}(0).
\end{equation}

In Eq.~\eqref{eq:hq_final_equation}, $\ket{p_0}$ and $\ket{\mathbf{1}}$ are
vectors belonging to the space of the classical configurations:
$\ket{\mathbf{1}}$ is a vector with all components set to $1$, while $\ket{p_0}
\equiv \mathbf{P}(0)$ is the initial probability distribution of the
configurations of the noise. In the case at hand, where we assume stationary
noise, all the configurations are equally probable and so
\begin{equation}
  \ket{p_0} = \frac 1 {N_c} \ket{\mathbf{1}}.
\end{equation}

The expression $\mel{\mathbf{1}}{A}{p_0} $ where $A$ is a $N_c (N_q^2-1) \times
N_c (N_q^2-1)$ matrix, denotes a partial inner product in the space of classical
configurations: the result is a $(N_q^2 -1)  \times (N_q^2 -1)$ matrix acting on
the Bloch vector of the quantum system.

Now let us focus on the study of continuous-time noisy quantum walk on the star
graph. If $N$ is the number of nodes in the graph, there are $N-1$ links and
thus $N-1$ independent RTN sources: the number of possible states of the noise
is $N_c=2^{N-1}$. The number of real parameters of the quantum system is $N^2 -
1$, and hence the number of rows of the matrix $\Hq$ is $2^{N-1}(N^2-1)$,
growing more than exponentially with $N$.

Evaluation of \eqref{eq:hq_final_equation} thus looks like a formidable task,
considering that matrix exponentiation is a costly function. However, the
matrices $V$ and $Q_E$ are largely sparse, with the number of nonzero elements
growing sub-exponentially with $N$: this allows us to resort to various
numerical techniques that ease the computational cost of
\eqref{eq:hq_final_equation}. While the matrix exponential of a sparse matrix is
dense, and thus its evaluation is still extremely costly, its action on a vector
$v$ can be evaluated just in terms of matrix-vector product operations
\cite{Sidje1998, AlMohy2011}.

By using the above techniques, we can evaluate a single exact dynamics for
multiple time instants for $N=10$ within seconds on a laptop. However, due to
the exponential scaling of the dimensions of $\Hq$, we cannot reach values of $N$
much higher than that, so this excludes, for example, the study of complete
graphs of more than $6$-$7$ nodes. Nevertheless, this method allows us to gain
insight into the dynamics of quantum walks affected by RTN on small graphs.
Further optimizations of the algorithm, using techniques of matrix compression
and distributed computation, may allow to reach even higher dimensions.

\section{Results}
\label{sec:results}
Both measures of non-Markovianity, defined in Eq.~\eqref{eqn:nonMgammaMax} and
Eq.~\eqref{eqn:BLP}, are highly sensitive to numerical errors in the evaluation
of the dynamics of the quantum walk, meaning that small fluctuations can lead to
completely wrong results (see the discussion in \cite{Benedetti2016}). Hence,
the need to employ the exact method presented in Sec.~\ref{sec:method}, instead
of the Montecarlo simulation used in \cite{Cattaneo2018}. Due to the numerical
complexity of the above method, we are restricted to a small number of RTN
sources. For this reason, we here consider quantum spatial search on the star
graph with central node as target, proven to be optimal in \cite{Cattaneo2018},
where it is also shown that the random telegraph noise with fast switching rate
$\mu$ has almost no effects on the probability of success of the search, while
decreasing $\mu$ leads to worse and worse results, proving that semi-static
noise jeopardizes the performance of the algorithm. Obviously, higher noise
strength $\nu$ implies lower success probability.

In this section we investigate if the presence of non-Markovianity is a resource
for quantum spatial search, i.e if it correlates with better performance of the
noisy algorithm. To do so, we employ both the measures of non-Markovianity
presented in Sec.~\ref{sec:nonM}.

\subsection{Non-Markovianity of the evolution according to the divisibility measure}

\begin{figure}[t] \includegraphics{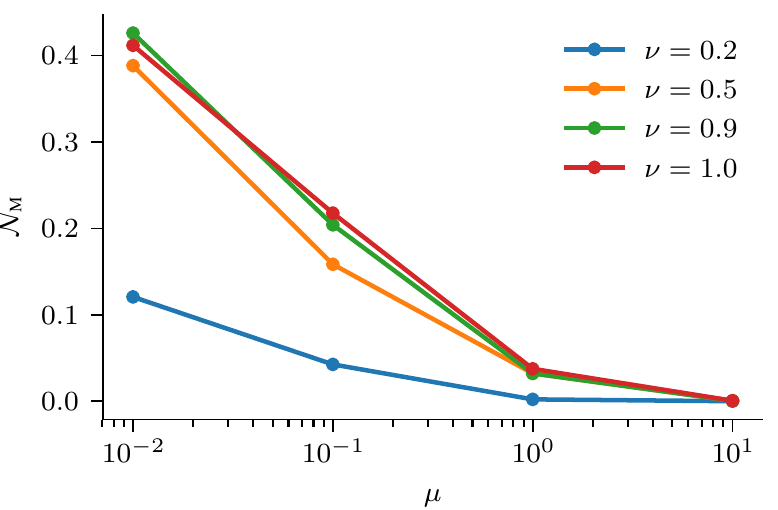}%
\caption{Divisibility measure of non-Markovianity $\nmdiv$ for the evolution of the initial state $\dyad{s}$ through the noisy algorithm of quantum spatial search, as a function of the switching rate $\mu$, for several values of the noise strength $\nu$. Non-Markovianity increases with the strength of the noise and decreases with the switching rate: strong, slow noise, which is the most detrimental, shows the greatest memory effects.}
\label{fig:nonMdivMeas}
\end{figure}

Considering the dynamics of the algorithm on a star graph with $N=7$ nodes and
central node as target, we have calculated $\Gamma(\tau,\tau_1)$ as defined in
Eq.~\eqref{eqn:nonMgammaTauTau1}, for the map defined in
Eq.~\eqref{eqn:evolution}, considering the starting state of the algorithm
$\rho(0) = \dyad{s}$. The maximum of $\Gamma(\tau, \tau_1)$ appears for finite
$\tau$ and $\tau_1$ because the dynamics has the maximally mixed state as fixed
point (as can be easily checked from Eq.~\eqref{eqn:evolution}). The actual
values for $\tau$ and $\tau_1$ vary with the parameters of the dynamics, but
accurate analysis has shown that, for $N \leq 10$, we can restrict to the region
$\tau, \tau_1 \leq 25$.

\begin{figure}
  \includegraphics{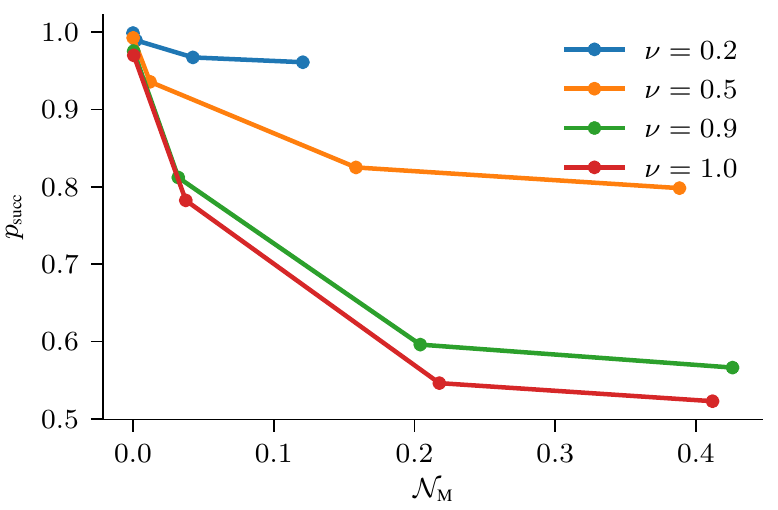}
  \caption{Success probability of the spatial search algorithm $p_\text{succ}$ 
  as a function of the non-Markovianity measure $\nmdiv$ of the dynamics, for 
  the same values of $\mu$ and $\nu$ of Fig.~\ref{fig:nonMdivMeas}. At fixed 
  noise strength, the success probability increases as the non-Markovianity decreases; 
  however, there is no clear correlation between $\nmdiv$ and $p_\text{succ}$.}
  \label{fig:success}
\end{figure}

The results for $\Gamma(\tau,\tau_1)$ are depicted in
Fig.~\ref{fig:nonMdivTauTau1}, for several values of $\mu$ and $\nu$.
Fig.~\ref{fig:nonMdivMeas} shows the value of the measure $\nmdiv$, obtained
after taking the maximum of all the values of $\Gamma(\tau,\tau_1)$ in
Fig.~\ref{fig:nonMdivTauTau1}. Apart from a slight bend from $\nu=0.9$ to
$\nu=1$ for $\mu=0.01$, we obtain higher values of $\nmdiv$ for slower and
stronger random telegraph noise, leading to bad performance of the algorithm.
Therefore, using such measure of non-Markovianity and in this specific case, the
presence of non-Markovianity is correlated with inefficient quantum spatial
search. 
In Fig. \ref{fig:success} we show the success
probability of the spatial search algorithm $p_\text{succ}$ as a function of the
non-Markovianity measure $\nmdiv$ of the dynamics, for the same values of $\mu$
and $\nu$ of Fig.~\ref{fig:nonMdivMeas}. At fixed noise strength, the success
probability increases as the non-Markovianity decreases. However, no clear
correlation between $\nmdiv$ and $p_\text{succ}$ may be seen.

\subsection{Non-Markovianity of the evolution according to the BLP measure}
To strengthen our results, we calculate the BLP measure of Eq.~\eqref{eqn:BLP}
as a second indicator of quantum non-Markovianity, again for the algorithm on a
star graph with $N=7$ and central node as target. The optimization over all the
possible initial states $\rho_1(0)$ and $\rho_2(0)$ is difficult to compute
efficiently, but for our purposes we just need to study the non-Markovianity of
the evolution of the CTQW, therefore we have kept fixed one of the two states,
say $\rho_1(0)$, as the initial state $\dyad{s}$, and we have optimized the
measure only over all the possible $\rho_2(0)$.

Numerical investigation showed that, keeping $\rho_1(0)=\dyad{s}$, we obtain the
maximum in Eq.~\eqref{eqn:BLP} by choosing as $\rho_2(0)$ the eigenstate
$\ket{r}$ of the Laplacian of the star graph, defined as:
\begin{equation}\label{eq:optimal_orthogonal_state}
  \ket{r}=-(N-1)\ket{1}+\sum_{k=2}^N\ket{k},
\end{equation}
where $N$ is the number of nodes in the graph and $\{\ket{k}\}_{k=1}^N$ is the
node basis.

\begin{figure}[t]
\includegraphics{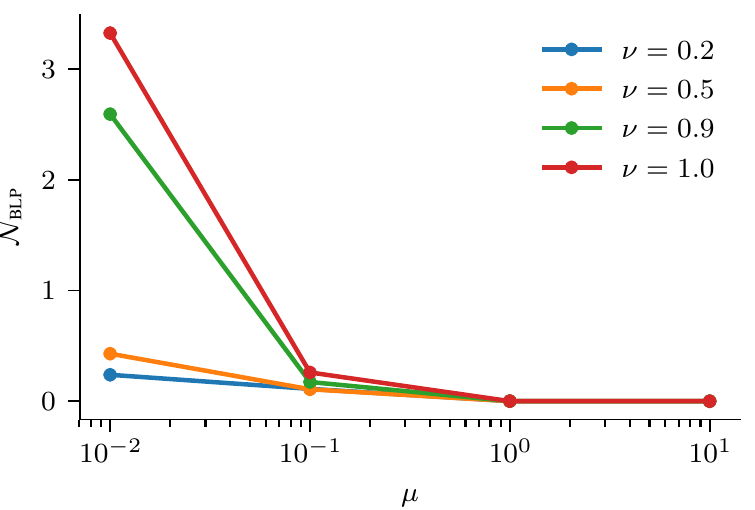}
\caption{BLP measure of non-Markovianity for the evolution of the initial state
 $\dyad{s}$ through the noisy algorithm of quantum spatial search, versus the
 switching rate $\mu$, for various values of the noise strength $\nu$. In
 computing the value of the BLP, we have considered as initial pair $\dyad{s}$
 and $\dyad{r}$, as defined in Eq.~\eqref{eq:optimal_orthogonal_state}. The
 presence of information backflow between system and environment is correlated
 with slow strong noise, i.e., with poorer performance of the algorithm. The
 stronger the noise the higher the non-Markovianity of the map. The measure is
 basically zero for switching rates above $\mu \simeq 1$, but the map is still
 non-Markovian, according to the divisibility measure $\nmdiv$.}
\label{fig:nonMBLP}
\end{figure}

The results for the BLP measure are shown in Fig.~\ref{fig:nonMBLP}, and they
perfectly confirm the correlation between presence of non-Markovianity in the
evolution and lower success probability of quantum spatial search.

Notice that this is one of the cases in which the divisibility measure proves to
be“higher” in the hierarchy of quantum non-Markovianity \cite{Li2018}. Indeed,
the divisibility measure detects the presence of non-Markovianity, although
small, for $\mu=10$ and $\mu=1$, while the BLP measure does not.

\subsection{Dependence on the size of the graph}

\begin{figure}
\includegraphics{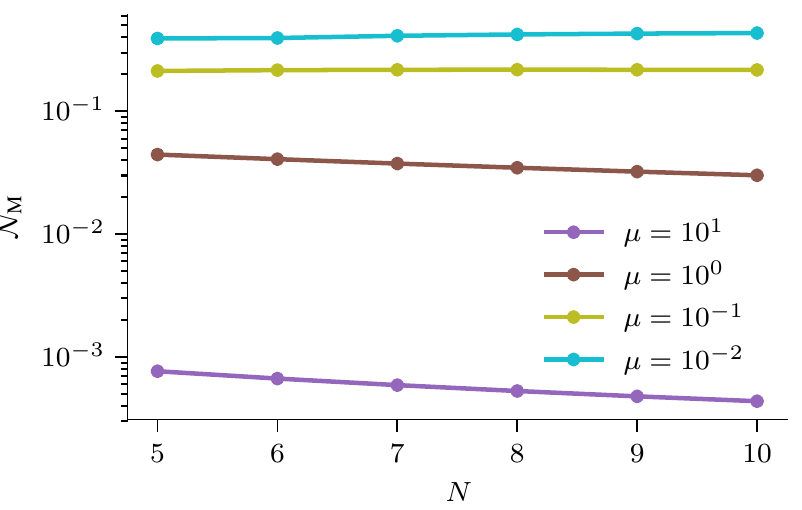}
\caption{Non-Markovianity measure $\nmdiv$ as a function of the size $N$ of the 
graph for different values of the switching rate $\mu$, for noise strength $\nu = 1.0$.
 The non-Markovianity measure slightly depends on $N$ (notice that the y axis is in 
 logarithmic scale), with $\nmdiv$ decreasing for fast noise, and basically constant 
 for slow noise. A qualitatively identical plot could be made for the BLP measure $\nmblp$.}
\label{fig:nonm_vs_N}
\end{figure}

The analysis above focused on the star graph with a central target node and
$N=7$. Here we address the dependence of non-Markovianity on the size of the
graph, by studying the two measures for different values of $N$, up to $N=10$,
so that the dynamics can be still evaluated with the exact method.

We found that the two quantities $\nmdiv$ and $\nmblp$ have a very similar
behavior as functions of $N$, and we show the former in
Fig.~\ref{fig:nonm_vs_N}, for different values of the switching rate and for the
maximum noise strength ($\nu=1$). We see that the non-Markovianity decreases
with $N$ for fast noise, while it is inappreciably increasing for slow noise.

While the computational complexity does not allow us to explore higher values of
$N$, we can expect non-Markovianity to maintain the same trend. This correlates
with the dependence of $p_\text{\tiny{succ}}$ on $N$, which is slightly
decreasing for strong, slow noise, and increasing for fast noise, as shown in
\cite{Cattaneo2018}, further confirming the link between non-Markovianity and
poorer performance of the algorithm.

%%%
\section{Concluding remarks}
\label{sec:concludingRemarks}
In this paper we have addressed spatial search implemented by CTQW on a star
graph and in the presence of RTN affecting the links between the nodes. In
particular, we have discussed the role of non-Markovianity of the quantum
dynamical map of the walker in determining the performance of the algorithm.
\par
In order to address the above problem, we have developed fast and optimized
code, not based on Montecarlo generation of stochastic trajectories, to achieve
a numerically exact solution of the dynamics of the walker. Avoiding
stochasticity allows one to increase the accuracy of the result and to reduce
fluctuations, a key requirement for evaluating most quantifiers of
non-Markovianity. The code is available online and can be applied to a general
quantum system affected by any number of RTN sources.
\par
Our results show that, unlike many other scenarios in which non-Markovianity can
be seen as a resource for various quantum information tasks, in the case at hand
spatial search performs better when the noise is fast, i.e., Markovian, as
opposed to slow noise, which induces a non-Markovian dynamics and is detrimental
for the algorithm. A possible intuitive explanation of the results above lies in
the fact that the typical recoherence effect due to the non-Markovianity of the
quantum map, happens on timescales that are much larger than the typical running
time of the algorithm. Notice also that there exists different physical
platforms in which state-of-the-art experiments are available with a
considerable dynamical control, and where this phenomena may be, in principle,
demonstrated.
\par
It is still unknown whether these conclusions are specific to the particular
statistics of the RTN, or if they are valid in a more general sense. Also, the
topology of the graph might play a role in the interplay between memory effects
and the localization of the walker in the target node. Further investigation
should hence address other graphs layouts, as well as other types of classical
or quantum noise that induce non-Markovian dynamics, and their effect on the
quantum spatial search algorithm.
\begin{acknowledgement}
This work has been supported by CARIPLO foundation through the Lake-of-Como
School program. MC has been supported by the EU through the Erasmus+ programme.
MACR and SM acknowledge support from the Academy of Finland via the Centre of
Excellence program (project 312058) and project 287750. MACR also acknowledges
financial support from the Turku Collegium for Science and Medicine. MGAP is
member of GNFM-INdAM. The authors are grateful to Claudia Benedetti, for useful
discussions.
\end{acknowledgement}

\bibliographystyle{apsrev4-1}
\bibliography{library}
\end{document}